# Unit cell determination of epitaxial thin films based on reciprocal space vectors by high-resolution X-ray diffractometry

*Dedicated to Prof. Dr. Johann Peisl on the Occasion of His 80th Birthday*

**Ping Yang[a]\*, Huajun Liu[b], Zuhuang Chen[c], Lang Chen[d] and John Wang[b]**

[a]Singapore Synchrotron Light Source (SSLS), National University of Singapore (NUS), 5 Research Link, 117603, Singapore, [b]Department of Materials Science and Engineering, National University of Singapore (NUS), 9 Engineering Drive 1, 117576, Singapore, [c]Department of Materials Science and Engineering and F. Seitz Materials Research Laboratory, University of Illinois Urbana-Champaign, Urbana, IL 61801, USA, and [d]South University of Science and Technology of China (SUSTC), Shenzhen, 518055, People's Republic of China

Correspondence email: slsyangp@nus.edu.sg

**Synopsis**

A new RSV (reciprocal space vector) method is developed, which provides a practical and concise route to crystal structure study of epitaxial thin films.

**Abstract**

A new approach, based on reciprocal space vectors (RSVs), is developed to determine Bravais lattice types and accurate lattice parameters of epitaxial thin films by high-resolution X-ray diffractometry (HR-XRD). The lattice parameters of single crystal substrates are employed as references to correct the systematic experimental errors of RSVs of thin films. The general procedure is summarized, involving correction of RSVs, derivation of raw unit cell, subsequent conversion to the Niggli unit cell and the Bravais unit cell by matrix calculation. Two methods of this procedure are described in 3D reciprocal space and 6D $G^6$-space, respectively. The estimation of standard error in the lattice parameters derived by this new approach is discussed. The whole approach is illustrated by examples of experimental data. The error of the best result is 0.0006 Å for the lattice parameter of ITO (Indium tin oxide) film. This new RSV method provides a practical and concise route to crystal structure study of epitaxial thin films, which could also be applied to the investigation of surface and interface structures.

## 1. Introduction

Epitaxial complex oxide thin films have attracted great attention due to their rich physical phenomena, such as ferroelectric, piezoelectric, ferromagnetic, multiferroic and superconducting, which promise novel functionalities in electronic devices (Zubko *et al.*, 2011). The physical properties of the epitaxial complex oxide thin films are very sensitive to the distortion of crystal structures, due to the strong coupling among charge, spin, orbital, and lattice (Schlom *et al.*, 2007, Hwang *et al.*, 2012, Dagotto, 2005). Thus, determination of crystal structures of the epitaxial thin films by high resolution X-ray diffractometry (HR-XRD), including the crystal symmetry and accurate lattice parameters, as well as domain structures and defects in films, is critical to understand and tune these physical properties (Cao & Cross, 1993).

Crystal symmetry has a significant effect on dielectric, ferroelectric and piezoelectric behavior of epitaxial ferroelectric/multiferroic thin films. The ionic displacement in the unit cell, which is determined by the crystal symmetry, gives rise to the ferroelectric polarization. For example, rhombohedral phase of $BaTiO_3$ (BTO) shows a ferroelectric polarization along the body diagonal direction <111>, while tetragonal phase along the *c*-axis direction <001> and orthorhombic phase along the face diagonal direction <110> (Kay & Vousden, 1949, Jaffe *et al.*, 1971). Piezoelectric property also strongly depends on the crystal symmetry. In the well-known $Pb(Zr,Ti)O_3$ (PZT) piezoelectric system, the highest piezoelectric response is observed near the morphotropic phase boundary (MPB) between a rhombohedral and a tetragonal phase (Jaffe *et al.*, 1971, Cao & Cross, 1993).

Lattice parameters of epitaxial thin films, which determine the unit cell distortion and film strain state, are important to tune such physical properties. For instance, the magnetic phase transition temperature of $LaCoO_3$ film increases with the increasing in-plane lattice parameters (Fuchs *et al.*, 2008). Another example is the doubling of superconducting transition temperature when in-plane lattice parameter of $La_{1.9}Sr_{0.1}CuO_4$ changes slightly from 3.78 to 3.76 Å (Locquet *et al.*, 1998). In $SrTiO_3$ (STO) thin film, tune the lattice parameter by substrate misfit strain can drive paraelectric phase into ferroelectric phase at room temperature (Haeni *et al.*, 2004).

In the determination of crystal structures, lattice and unit cell dimension are always the first step to start, i.e., to begin with Bravais lattice type and lattice parameters. For poly-crystalline film, with powder diffractometry and related computational algorithms (Pecharsky & Zavalij, 2003), its diffraction pattern can be indexed and the lattice parameters can be obtained with satisfactory precision. Some software is *ab initio* indexing, such as TREOR & DICVOL employing trial-and-error method, and ITO zone searching combined with Delaunay-Ito technique, pp. 399 (Pecharsky & Zavalij, 2003). A simplified diagram for both bulk and film samples is as shown in the left-half of Fig. 1, which outlines a general procedure on the determination of crystal structures, for either crystalline or poly-crystalline samples.

For a single crystalline epitaxial film, above procedure could in principle be used, for example, in some single crystal structure determination software packages, such as SMART APEX II, Bruker AXS (APEX2, 2009). About 25 reflections (reciprocal space points) could be used to deduce lattice parameters, as in normal case of single crystal structure determination. However, it often does not work for a film on substrate due to following facts that

1) Two sets of diffraction spots (from film and substrate respectively) cause an inconsistency in the diffraction pattern and, mess up the procedure and result in no lattice parameters deduced;

2) Weaker diffraction intensity from film yields fewer useful reflections and less-precise peak positions. Such diffraction peak usually has broader width and even extended Bond methods offers unsatisfactory peak position (Bond, 1960, Schmidbauer *et al.*, 2012). In the experiments, typical accuracy in Bragg angle $\theta_B$ is in the order of 0.002 °≅$3.5 \times 10^{-5}$ rad. for a bulk sample with a HR-XRD set-up. But for a film peak, the accuracy is much worsened (Schmidbauer *et al.*, 2012). For example, about 0.01 ° deviation in Bragg angle $\theta_B$ can lead an error as larger as 0.2 ° than $\beta$=90 ° to the monoclinic angle for a tetragonal lattice of BTO at 130 ℃ (para-electric to ferro-electric phase transition point), calculated from *002*, *013* and *103* diffraction peaks (Yang, October 2012). It is hard to judge to which crystallographic system, tetragonal or monoclinic, it belongs. Such error in the monoclinic angle is crucial to recognize the crystal system and symmetry at the MPB of some multi-ferroic films as above mentioned (Zeches *et al.*, 2009);

3) As X-ray beam cannot penetrate through the substrate in normal case of the wavelength, back-reflection geometry has to be adopted in most experiments. Limited number of diffraction peaks from film can be collected, again, reducing a chance to obtain accurate lattice parameters;

4) Existence of twining variants results in wrong peak positioning, leading to wrong lattice parameters and even failing to do so, due to widening and shifting of the peak positions.

In these methods, inaccurate Bragg angles $\theta_B$ and fewer numbers of diffraction peaks make it hard to obtain accurate lattice parameters and weaker diffraction intensities make the Bond methods less effective. More seriously, only *d*-spacing (either accurate or less) could be obtained at first, from which it is hard to deduce the Bravais lattice type from fewer Bragg planes. "Guess and check" method may often be used in this case, relying mostly on one's luck during the work.

For the procedure after "Bravais lattice" in Fig. 1, we note that a new method to solve and refine the crystal structure (atomic positions inside the crystal cell) has been demonstrated recently for CuMnAs film on a GaAs substrate (Wadley *et al.*, 2013).

In this article, we focus on a reciprocal space vector (RSV) method using HR-XRD set-up, in order to overcome the above-listed difficulties in lattice parameter determination for single crystalline epitaxial film, which obtains firstly the Niggli basis vectors accurately with the lattice parameters of the substrate as a reference. The Bravais lattice type together with the lattice parameters is then worked out. In this RSV method, relationship between the lattices of film and those of substrate is clearly revealed and the lattice parameters can be obtained with high accuracy. The procedure is as shown in the right-half of Fig. 1.

**2. Method 1: Reciprocal space and RSVs**

In epitaxial film characterization, 2-dimensional (2D) reciprocal space mapping (RSM) has been widely used to obtain necessary RSVs or Bragg angles $\theta_B$ for the determination of the lattice parameters (Catalan et al., 2007, Chu et al., 2009, Daumont et al., 2010, Liu, Yao, et al., 2010, Qi et al., 2005, Noheda et al., 2002, Bai et al., 2004, Liu, Yang, et al., 2010, Saito et al., 2006). There is always a pre-set assumption that the crystal axes of film are in the RSM plane. In other words, the angle between the crystal axes in-film-plane are pre-assumed in this method, to be the mutual angle between the RSMs.

In this section, RSV method in 3-dimensional (3D) reciprocal space is introduced, showing a more accurate determination of crystal lattice parameters than RSM without any pre-set assumption. This method starts with measurement of film RSVs, which is then corrected by rescaling and rotating referred to the substrate RSVs. The Niggli cell is reduced and the Bravais lattice type with the lattice parameters are finally worked out.

**2.1. Obtaining basis vectors in reciprocal and real space**

In 3D reciprocal space, an RSV can generally be represented as (*HKL*) or matrix $\begin{pmatrix} H \\ K \\ L \end{pmatrix}$, i.e.,

$$\boldsymbol{q} = H\boldsymbol{a}^* + K\boldsymbol{b}^* + L\boldsymbol{c}^* = (HKL) = \begin{pmatrix} H \\ K \\ L \end{pmatrix} \qquad (1)$$

where *H*, *K* & *L* are the coordinates based on selected bases in the reciprocal space along **H**, **K** & **L** directions, respectively. A coordinate system for the reciprocal space with basis vectors **a***, **b*** & **c*** shown in Fig. 2. *H*, *K* & *L* are integers (Miller indices) if **a***, **b*** & **c*** are chosen from the reciprocals of its own basis vectors **a**, **b** & **c** in real space and RSV falls on its lattice point, i.e., $\boldsymbol{q_0}$; *H*, *K* & *L* may not be integers if other set of the basis vectors $\boldsymbol{a}^*$, $\boldsymbol{b}^*$ & $\boldsymbol{c}^*$ are chosen, for example, from its substrate lattice parameters, or when it does not fall on the lattice point, i.e., $\boldsymbol{q}$, in general. Note that there is another definition of the RSV and the scattering vector (below): $\boldsymbol{Q}=2\pi\boldsymbol{q}$. For simplicity we discuss $\boldsymbol{q}$ here.

By properly selecting three RSVs, e.g., (*00L*), (*H0L*) and (*0KL*), three basis vectors, $\boldsymbol{a}^*$, $\boldsymbol{b}^*$ & $\boldsymbol{c}^*$ can be obtained as follows

$$\begin{cases} \boldsymbol{a}^* = \frac{1}{H}[(H0L) - (00L)] \\ \boldsymbol{b}^* = \frac{1}{K}[(0KL) - (00L)] \\ \boldsymbol{c}^* = \frac{1}{L}(00L) \end{cases} \quad (2a)$$

or written as

$$\begin{cases} \boldsymbol{a}^* = \frac{1}{H}\left[\begin{pmatrix} H \\ 0 \\ L \end{pmatrix} - \begin{pmatrix} 0 \\ 0 \\ L \end{pmatrix}\right] \\ \boldsymbol{b}^* = \frac{1}{K}\left[\begin{pmatrix} 0 \\ K \\ L \end{pmatrix} - \begin{pmatrix} 0 \\ 0 \\ L \end{pmatrix}\right] \\ \boldsymbol{c}^* = \frac{1}{L}\begin{pmatrix} 0 \\ 0 \\ L \end{pmatrix} \end{cases}. \quad (2b)$$

So-obtained bases in reciprocal space may not be primitive and they may also have systematic errors from the measurement in the experiment and methodology. They can be rectified by making corrections to the RSVs with the substrate as a reference in next section.

## 2.2. Correction of RSVs using substrate as a reference

As measured RSVs can be different in their lengths and orientations from their correct RSVs in reciprocal space, the correction should be made and classified as scaling part $s$ (*scale factor*) and rotational part $\boldsymbol{R}$ (*rotation matrix*). Such systematic errors may be from incorrect wavelength and inaccurate angle measurement in the experiment as discussed above. Scale factor $s$ denotes the length ratio of correct RSV to its measured RSV. Rotation matrix $\boldsymbol{R}$ can be decomposed into two rotations about *H*-axis and *K*-axis respectively, see pp. 76 (Giacovazzo *et al.*, 2002), if a right-handed Cartesian coordinate system *H, K* & *L* is set up (in Fig. 2). In other cases, non-Cartesian coordinate system could be converted into Cartesian coordinate system, and then do the rotation correction, pp. 74 (Giacovazzo *et al.*, 2002).

As RSVs of film are usually in the vicinities of those of substrate, we have a reason to conclude that such systematic errors arising in the measurement for the film should be the same or very close to those for the substrate. The basic idea is to find a correction for substrate RSVs first, then make the same correction to film RSVs.

The corrected RSV ($H_1K_1L_1$) is related to measured RSV ($h_1k_1l_1$) of the substrate (shorted as *sub*)

by
$$\begin{pmatrix} H_1 \\ K_1 \\ L_1 \end{pmatrix}_{\substack{corrected \\ for\ sub}} = s\boldsymbol{R} \begin{pmatrix} h_1 \\ k_1 \\ l_1 \end{pmatrix}_{\substack{measured \\ for\ sub}}. \tag{3a}$$

As corrected RSV ($H_1K_1L_1$) of substrate is known, $s$ and $\boldsymbol{R}$ can be derived from Eq. (3a). Then we can make the same correction to film RSVs,

$$\begin{pmatrix} H_2 \\ K_2 \\ L_2 \end{pmatrix}_{\substack{corrected \\ for\ film}} = s\boldsymbol{R} \begin{pmatrix} h_2 \\ k_2 \\ l_2 \end{pmatrix}_{\substack{measured \\ for\ film}}. \tag{3b}$$

Scale factor $s$ is a length ratio of the RSV ($H_1K_1L_1$) to RSV ($h_1k_1l_1$) for substrate as in Eq. (3a).

$\boldsymbol{R}$ can be represented as

$$\boldsymbol{R} = \boldsymbol{R}_H(\alpha) \cdot \boldsymbol{R}_K(\beta) \tag{4a}$$

$$\boldsymbol{R}_H(\alpha) = \begin{pmatrix} 1 & 0 & 0 \\ 0 & \cos\alpha & -\sin\alpha \\ 0 & \sin\alpha & \cos\alpha \end{pmatrix} \tag{4b}$$

$$\boldsymbol{R}_K(\beta) = \begin{pmatrix} \cos\beta & 0 & \sin\beta \\ 0 & 1 & 0 \\ -\sin\beta & 0 & \cos\beta \end{pmatrix} \tag{4c}$$

where $\alpha$ and $\beta$ are the rotating angles of ($h_1k_1l_1$) about *H*-axis and *K*-axis respectively to get coincident with correct ($H_1K_1L_1$) for the substrate, i.e., rotating correction. Counter-clockwise rotation is positive.

Example for correction of RSV (013): (*-0.0007  0.9996  3.0006*) and (*-0.0011  1.0065  2.9002*) were obtained for SrTiO$_3$ (STO) substrate and BiFeO$_3$ (BFO) film respectively. A correction with $s$=0.9999 and $R_H$(2.333×10$^{-4}$), $R_K$(-1.800×10$^{-4}$), leads it to (*013*) for the substrate as it should be, with the rotation angles 0.01337 °about *H*-axis and -0.01031 °about *K*-axis respectively. The RSV for BFO film is accordingly corrected as (*-0.0004  1.0069  2.8996*) using the same scale factor $s$ and rotation matrix $\boldsymbol{R}$.

### 2.3. Basis vectors of film in real space

Using Eqs. (2) with corrected RSVs from Eqs. (3), a raw and primitive unit cell with the shortest vectors *a*, *b* & *c* of film in real space can thus be derived as

$$\begin{cases} \boldsymbol{a} = \dfrac{\boldsymbol{b}^* \times \boldsymbol{c}^*}{V^*} \\ \boldsymbol{b} = \dfrac{\boldsymbol{c}^* \times \boldsymbol{a}^*}{V^*} \\ \boldsymbol{c} = \dfrac{\boldsymbol{a}^* \times \boldsymbol{b}^*}{V^*} \end{cases} \quad (5)$$

$$V^* = \frac{1}{V} = \boldsymbol{a}^* \times \boldsymbol{b}^* \cdot \boldsymbol{c}^*. \quad (6)$$

where $V^*$ and $V$ are the volumes of the raw cell in reciprocal and real space respectively.

### 2.4. Niggli cell and Bravais lattice of film

Raw bases *a*, *b* & *c* of film can be derived from Eq. (5). It can be reduced to a standard Niggli cell which should satisfy the conditions for two types that

$$\boldsymbol{a} \cdot \boldsymbol{a} \leq \boldsymbol{b} \cdot \boldsymbol{b} \leq \boldsymbol{c} \cdot \boldsymbol{c} \text{ and } T = (\boldsymbol{a} \cdot \boldsymbol{b})(\boldsymbol{b} \cdot \boldsymbol{c})(\boldsymbol{c} \cdot \boldsymbol{a}) \quad (7)$$

Type I:    $T>0$    and    $\boldsymbol{a} \cdot \boldsymbol{b} > 0,\ \boldsymbol{b} \cdot \boldsymbol{c} > 0, \boldsymbol{c} \cdot \boldsymbol{a} > 0;$    (8)

Type II:   $T \leq 0$   and    $\boldsymbol{a} \cdot \boldsymbol{b} \leq 0,\ \boldsymbol{b} \cdot \boldsymbol{c} \leq 0, \boldsymbol{c} \cdot \boldsymbol{a} \leq 0.$   (9)

Other main conditions and special conditions can be found in Chapter 9.2, pp. 750 (Hahn, 2006) or pp. 19 (Ma Zhesheng & Shi Nicheng, 1995).

Using Table 4 (Andrews & Bernstein, 1988) or Table 1 (Paciorek & Bonin, 1992), the raw cell can be transformed into a standard Niggli cell by multiplying the transformation matrix. It is actually a projection of the raw cell to a subspace (to the standard type of Niggli cell) in $G^6$ space, as described in next section Eq. (18).

Using Table 9.2.5.1., Chapter 9.2, pp. 753 (Hahn, 2006), conventional Bravais lattice type and the lattice parameters can be derived from the standard Niggli cell using its transformation matrix $\boldsymbol{M}_{N \to B}$ from the table that

$$\begin{pmatrix} a \\ b \\ c \end{pmatrix}_{Bravais} = \boldsymbol{M}_{N \to B} \begin{pmatrix} a \\ b \\ c \end{pmatrix}_{Niggli}. \quad (10)$$

For example, the transformation matrix for Bravais lattice type monoclinic *mC* (No. 10) is

$$\boldsymbol{M}_{N \to B} = \begin{pmatrix} 1 & 1 & 0 \\ 1 & -1 & 0 \\ 0 & 0 & -1 \end{pmatrix}. \quad (11)$$

<u>Example of determining a Bravais lattice:</u> as partly shown in the example in section **2.2.**, other RSVs of the same sample are corrected with similar procedure using Eq. (3) to be (*002*) and ($\bar{1}03$), respectively, from (*0.0000 0.0000 1.9992*) and (*-0.9996 0.0000 2.9993*) for the substrate; and to be (*0.0000 0.0034 1.9382*) and (*-1.0004 0.0080 2.9009*), from measured (*0.0000 0.0034 1.9374*) and

(*-1.0000  0.0080  2.9002*) for the film. Such correction leads to a raw cell that $a$=3.898 Å, $b$=3.904 Å & $c$=4.030 Å, $\alpha$=90.36 °, $\beta$=90.34 ° & $\gamma$=90.19 °. It is the case of reduced bases No. 10, monoclinic *mC* lattice, Table 9.2.5.1. (Hahn, 2006). From Table 4 (Andrews & Bernstein, 1988) or Table 1 (Paciorek & Bonin, 1992), the raw cell can be reduced into a standard Niggli cell that $a=b$=3.901 Å, & $c$=4.030 Å, $\alpha=\beta$=90.35 ° & $\gamma$=90.19 °. The Bravais lattice parameters have finally been derived that $a$=5.507 Å, $b$=5.526 Å & $c$=4.030 Å, $\alpha$=90 °, $\beta$=90.50 ° & $\gamma$=90 °, using the transformation matrix No. 10, Table 9.2.5.1. (Hahn, 2006), also as represented in Eq. (11).

Such transformations can be performed using an *ACCEL* calculation program *DeFLaP* (Determination of Film Lattice Parameters) developed by the authors. As one of the directions of incident beam owns poor resolution, $\gamma$ angle has bigger error than others, which will be discussed in sections **3.3.** and **4.** below.

## 3. Method 2: Vectors in $G^6$ space and unit cell

In this section, a vector in $G^6$ space is treated to show how they represent a unit cell. Similar correction can be made with substrate as a reference, but simpler, without separating of rescaling and rotating parts as above. The basis vectors of Niggli cell are then derived and Bravais lattice type is determined with the projection method as used in last section.

### 3.1. Representation of a unit cell in G6 space

A unit cell can be represented as a vector (point) in a 6-dimension (6D) Euclidean space, denoted as $G^6$ space (Andrews & Bernstein, 1988). Actually, from the metric matrix

$$\begin{pmatrix} \bm{a}\cdot\bm{a} & \bm{a}\cdot\bm{b} & \bm{a}\cdot\bm{c} \\ \bm{b}\cdot\bm{a} & \bm{b}\cdot\bm{b} & \bm{b}\cdot\bm{c} \\ \bm{c}\cdot\bm{a} & \bm{c}\cdot\bm{b} & \bm{c}\cdot\bm{c} \end{pmatrix} \tag{12}$$

there are only 6 independent components, forming a so-called Niggli matrix (Niggli, 1928)

$$\begin{pmatrix} \bm{a}\cdot\bm{a} & \bm{b}\cdot\bm{b} & \bm{c}\cdot\bm{c} \\ \bm{b}\cdot\bm{c} & \bm{a}\cdot\bm{c} & \bm{a}\cdot\bm{b} \end{pmatrix}. \tag{13}$$

It forms a vector $\bm{g}$ in $G^6$ space

$$\bm{g} = \begin{pmatrix} g_1 \\ g_2 \\ g_3 \\ g_4 \\ g_5 \\ g_6 \end{pmatrix} = \begin{pmatrix} \bm{a}\cdot\bm{a} \\ \bm{b}\cdot\bm{b} \\ \bm{c}\cdot\bm{c} \\ 2\bm{b}\cdot\bm{c} \\ 2\bm{c}\cdot\bm{a} \\ 2\bm{a}\cdot\bm{b} \end{pmatrix} = \begin{pmatrix} a^2 \\ b^2 \\ c^2 \\ 2bc\cdot\cos\alpha \\ 2ac\cdot\cos\beta \\ 2ab\cdot\cos\gamma \end{pmatrix}. \tag{14}$$

Any such one vector in $G^6$ space corresponds uniquely to one unit cell with the lattice parameters $a$, $b$ & $c$, $\alpha$, $\beta$ & $\gamma$. It brings a convenience to derive lattice parameters by using this representation in $G^6$ space. A correction of the lattice parameters of a unit cell becomes simpler and described below.

### 3.2. Correction of film lattice parameters in G6 space

In a real experiment, lattice parameters can be obtained, for example, from Eq. (5) to convert it to its real space parameters for a substrate and film respectively. There are unavoidably systematic errors in the measurement as discussed in section **2.2.** If a correction vector $\Delta \boldsymbol{g}$ in $G^6$ space is needed for substrate to obtain its standard lattice parameters, following equations should be obtained and $\Delta \boldsymbol{g}$ can be derived that

$$\boldsymbol{g}_{sub} = \begin{pmatrix} a^2 \\ b^2 \\ c^2 \\ 2bc \cdot \cos\alpha \\ 2ac \cdot \cos\beta \\ 2ab \cdot \cos\gamma \end{pmatrix}_{\substack{standard \\ for\ sub}} = \Delta \boldsymbol{g} + \begin{pmatrix} a^2 \\ b^2 \\ c^2 \\ 2bc \cdot \cos\alpha \\ 2ac \cdot \cos\beta \\ 2ab \cdot \cos\gamma \end{pmatrix}_{\substack{measured \\ for\ sub}} \quad (15)$$

where the subscript *sub* means *substrate*. The correction vector $\Delta \boldsymbol{g}$ is written as

$$\Delta \boldsymbol{g} = \begin{pmatrix} \Delta g_1 \\ \Delta g_2 \\ \Delta g_3 \\ \Delta g_4 \\ \Delta g_5 \\ \Delta g_6 \end{pmatrix}. \quad (16)$$

As the corresponding vectors in $G^6$ space for substrate and film are close, the systematic error caused in an experiment should apparently be the same or close as stated in section **2.2.** Then the same correction should be made to the vector $\boldsymbol{g}_{raw\ film}$ in $G^6$ space for the film as for the substrate that

$$\boldsymbol{g}_{raw\ film} = \begin{pmatrix} g_1 \\ g_2 \\ g_3 \\ g_4 \\ g_5 \\ g_6 \end{pmatrix}_{\substack{raw \\ corrected}} = \begin{pmatrix} a^2 \\ b^2 \\ c^2 \\ 2bc \cdot \cos\alpha \\ 2ac \cdot \cos\beta \\ 2ab \cdot \cos\gamma \end{pmatrix}_{\substack{raw \\ corrected}} = \Delta \boldsymbol{g} + \begin{pmatrix} a^2 \\ b^2 \\ c^2 \\ 2bc \cdot \cos\alpha \\ 2ac \cdot \cos\beta \\ 2ab \cdot \cos\gamma \end{pmatrix}_{\substack{measured \\ for\ film}}. \quad (17)$$

Through above calculation, raw lattice parameters of film can be derived with correction $\Delta \boldsymbol{g}$ from the measured lattice parameters, to its substrate as a reference. This raw cell can be converted to a standard Niggli cell and Bravais lattice will be finally determined with Eq. (10) and shown below.

Using Table 4 (Andrews & Bernstein, 1988) or Table 1 (Paciorek & Bonin, 1992), a raw cell corrected in Eq. (17) can be transformed (projected) into a *Niggli-reduced cell* by

$$\boldsymbol{g}_{Niggli-reduced} = \begin{pmatrix} a^2 \\ b^2 \\ c^2 \\ 2bc \cdot \cos\alpha \\ 2ac \cdot \cos\beta \\ 2ab \cdot \cos\gamma \end{pmatrix}_{Niggli-reduced} = \boldsymbol{M}_{R \to N} \cdot \boldsymbol{g}^{raw}_{film} = \boldsymbol{M}_{R \to N} \begin{pmatrix} a^2 \\ b^2 \\ c^2 \\ 2bc \cdot \cos\alpha \\ 2ac \cdot \cos\beta \\ 2ab \cdot \cos\gamma \end{pmatrix}_{raw\ corrected} \quad (18)$$

where the raw cell is projected as a Niggli-reduced cell (onto a subspace in $G^6$ space), $\boldsymbol{M}_{R \to N}$ the projection matrix from Table 4 (Andrews & Bernstein, 1988) or Table 1 (Paciorek & Bonin, 1992). The Bravais lattice will be subsequently determined using the transformation matrix $\boldsymbol{M}_{N \to B}$ from Eq. (10).

### 3.3. Error estimation

The lattice parameters *a, b* & *c*, *α, β* & *γ* of the Niggli cell can generally be expressed by the projected vector (Niggli-reduced cell) with an error vector $\delta\boldsymbol{g}$ in $G^6$ space, i.e.,

$$\begin{pmatrix} a^2 \\ b^2 \\ c^2 \\ 2bc \cdot \cos\alpha \\ 2ac \cdot \cos\beta \\ 2ab \cdot \cos\gamma \end{pmatrix} = \begin{pmatrix} a^2 \\ b^2 \\ c^2 \\ 2bc \cdot \cos\alpha \\ 2ac \cdot \cos\beta \\ 2ab \cdot \cos\gamma \end{pmatrix}_{Niggli-reduced} \pm \delta\boldsymbol{g} \quad . \quad (19)$$

The error or deviation of reduced Niggli cell, $\delta\boldsymbol{g}$, can be calculated using the distance between the projected vector (Niggli-reduced cell, Eq. (18)) and its raw vector (raw corrected cell, Eq. (17)) in $G^6$ space:

$$\delta\boldsymbol{g} = \begin{pmatrix} a^2 \\ b^2 \\ c^2 \\ 2bc \cdot \cos\alpha \\ 2ac \cdot \cos\beta \\ 2ab \cdot \cos\gamma \end{pmatrix}_{Niggli-reduced} - \begin{pmatrix} a^2 \\ b^2 \\ c^2 \\ 2bc \cdot \cos\alpha \\ 2ac \cdot \cos\beta \\ 2ab \cdot \cos\gamma \end{pmatrix}_{raw\ corrected} \quad . \quad (20)$$

Or according to Eq. (18), we have

$$\delta \boldsymbol{g} = (\boldsymbol{M}_{R \to N} - E) \cdot \begin{pmatrix} a^2 \\ b^2 \\ c^2 \\ 2bc \cdot \cos \alpha \\ 2ac \cdot \cos \beta \\ 2ab \cdot \cos \gamma \end{pmatrix}_{\substack{raw \\ corrected}} \quad (21)$$

where E is a unit matrix.

Example 1 to estimate errors:

For the Niggli-reduced cell of BFO film as shown in the examples in section **2.2** and **2.4**, we can calculate the deviation as follows

$$\delta \boldsymbol{g} = \begin{pmatrix} 3.901^2 \\ 3.901^2 \\ 4.030^2 \\ 2 \times 3.901 \times 4.030 \times \cos 90.35° \\ 2 \times 3.901 \times 4.030 \times \cos 90.35° \\ 2 \times 3.901 \times 3.901 \times \cos 90.19° \end{pmatrix}_{\substack{Niggli- \\ reduced}} - \begin{pmatrix} 3.898^2 \\ 3.904^2 \\ 4.030^2 \\ 2 \times 3.904 \times 4.030 \times \cos 90.36° \\ 2 \times 3.898 \times 4.030 \times \cos 90.34° \\ 2 \times 3.898 \times 3.903 \times \cos 90.19° \end{pmatrix}_{\substack{raw \\ corrected}}$$

$$= \begin{pmatrix} 0.021 \\ -0.021 \\ 0.000 \\ 0.008 \\ -0.008 \\ 0.000 \end{pmatrix}. \quad (22)$$

From the first two components in Eq. (22), using Eq. (19) we have

$$2 \cdot (a^2 - 3.901^2)^2 = 0.021^2 + (-0.021)^2$$

$$a^2 - 3.901^2 = \pm 0.021$$

$$a = 3.901 \pm 0.003 \text{ (Å)}.$$

From the fourth and fifth components in Eq. (22), using Eq. (19) we have

$$(2bc \cdot \cos \alpha - 2 \times 3.901 \times 4.030 \times \cos 90.35°)^2 +$$
$$(2ac \cdot \cos \beta - 2 \times 3.901 \times 4.030 \times \cos 90.35°)^2 = 0.008^2 + (-0.008)^2.$$

As the reduced cell has $a=b=3.901$ Å, & $c=4.030$ Å, $\alpha=\beta=90.35°$, & $\gamma=90.19°$,

$$2 \times [2 \times 3.901 \times 4.030 \times (\cos \beta - \cos 90.35°)]^2 = 2 \times 0.008^2$$

$$2 \times 3.901 \times 4.030 \times (\cos \beta - \cos 90.35°) = \pm 0.008$$

$$-2 \times 3.901 \times 4.030 \times \sin 90.35° \times \text{Radians}(\beta - 90.35°) = \pm 0.008$$

$$\beta = 90.35° \pm 0.015°.$$

To show the procedure, the significant figure of the numbers is ignored in above calculation.

Example 2 to estimate errors:

To reduce the Niggli cell of an ITO film (tin-doped indium oxide), raw lattice parameters were obtained from RSVs of the film that $a$=5.0683 Å, $b$=5.0698 Å and $c$=5.0694 Å, $\alpha$=89.994°, $\beta$=89.958° & $\gamma$=89.992°. It is the case of reduced bases No. 3, cubic $cP$ lattice (Table 9.2.5.1. (Hahn, 2006). From Table 4 (Andrews & Bernstein, 1988) or Table 1 (Paciorek & Bonin, 1992), the Niggli-reduced cell was projected that $a=b=c$=5.0691 Å, $\alpha=\beta=\gamma$=90°. The deviation is calculated to be

$$\delta \boldsymbol{g} = \begin{pmatrix} 5.0691^2 \\ 5.0691^2 \\ 5.0691^2 \\ 0 \\ 0 \\ 0 \end{pmatrix}_{\substack{Niggli-\\reduced}} - \begin{pmatrix} 5.0683^2 \\ 5.0698^2 \\ 5.0694^2 \\ 2 \times 5.0698 \times 5.0694 \times \cos 89.994° \\ 2 \times 5.0683 \times 5.0694 \times \cos 89.958° \\ 2 \times 5.0683 \times 5.0698 \times \cos 89.992° \end{pmatrix}_{\substack{raw\\corrected}} = \begin{pmatrix} 0.0090 \\ -0.0062 \\ -0.0027 \\ -0.0050 \\ -0.0374 \\ -0.0073 \end{pmatrix}. \quad (23)$$

Using Eq. (19), a cubic cell can then be expressed as

$$\begin{pmatrix} a^2 \\ a^2 \\ a^2 \\ 2a^2 \cdot \cos \alpha \\ 2a^2 \cdot \cos \beta \\ 2a^2 \cdot \cos \gamma \end{pmatrix} = \begin{pmatrix} 5.0691^2 \\ 5.0691^2 \\ 5.0691^2 \\ 0 \\ 0 \\ 0 \end{pmatrix} \pm \begin{pmatrix} 0.0090 \\ -0.0062 \\ -0.0027 \\ -0.0050 \\ -0.0374 \\ -0.0073 \end{pmatrix}. \quad (24)$$

From the first 3 components, we have

$$3 \cdot (a^2 - 5.0691^2)^2 = 0.0090^2 + (-0.0062)^2 + (-0.0027)^2 = (0.0112)^2.$$

Hence, $a$=5.0691±0.0006 (Å) for the cubic ITO film.

## 4. Experimental condition and resolution

Using a diffractometer, angles or diffraction positions for a diffraction peak can be exactly measured, through which corresponding scattering vector $\boldsymbol{q}$ or RSV coordinates can be obtained.

### 4.1. A scattering vector in four-circle diffractometer coordinate systems

A standard four-circle diffractometer is used, for example, and the coordinate systems are as shown in Fig. 3. The coordinate system convention is as proposed in the SPEC, pp. 163 (Certified Scientific Software, 2008). Three orthogonal and right-handed coordinate systems are established, i.e.,

1) laboratory coordinate system (fixed frame in laboratory, Fig. 3(a)); 2) diffractometer coordinate system (angular as shown in Fig. 3(a) with Euler circles $2\theta$, $\omega$, $\chi$ and $\phi$) and 3) sample coordinate system (fixed with sample natural axes, as shown in Fig. 3(b) with a scattering vector $\boldsymbol{q}$. Circles $2\theta$, $\omega$, and $\phi$, are defined as right-handed and $\chi$ left-handed. The other definition of the coordinate systems and rotations can be found in (Busing & Levy, 1967).

As mentioned in section 2.1., a scattering vector $\boldsymbol{Q}=\boldsymbol{K}_2-\boldsymbol{K}_1$ or $\boldsymbol{Q}=2\pi\boldsymbol{q}$. $\boldsymbol{q}$ is represented as

$$\boldsymbol{q} = \boldsymbol{k}_2 - \boldsymbol{k}_1 \qquad (25)$$

and oriented with $\chi$ and $\phi$ angles as shown in Fig. 3(b). $\boldsymbol{K}_1$ is the wave vector of incident X-rays and $\boldsymbol{K}_2$ of scattered X-rays ($|\boldsymbol{K}_1|=|\boldsymbol{K}_2|=2\pi/\lambda$). Correspondingly, $\boldsymbol{k}_1$ is the wave number vector of incident X-rays and $\boldsymbol{k}_2$ of scattered X-rays as shown in Fig. 3 ($|\boldsymbol{k}_1|=|\boldsymbol{k}_2|=1/\lambda$). $\boldsymbol{q}$ has component $q_z$ along $z$-direction and $q_\phi$ in the $xy$-plane. If it is along the normal direction of Bragg planes and satisfied with a Bragg condition, its magnitude $q$ is then equal to a reciprocal space vector – RSV and diffraction then occurs.

A case is as shown in Fig. 3(c) that the scattering vector $\boldsymbol{q}$ is firstly rotated into the scattering plane, i.e., $xy$-plane in the coordinate system with angle $\theta$-$\omega$ against the $q_z$ axis and it is then rotated to the Bragg condition. For example, BFO (103), whole crystal can be rotated -90° by $\chi$. $q_z$ now is along $x$-direction. It is then rotated another 90° by $\phi$ (i.e. by $q_z$) and (103) is in $xy$-plane at an angle against $q_z$. With a subsequent rotation $\omega$ in the scattering plane, one then gets diffraction. The incident X-ray is at angle $\alpha$ ($\equiv\omega$) and diffracted X-rays at angle $\beta$ respectively to the component $q_\phi$. The magnitude $q$ is worked out as

$$q = 2k \cdot \sin\frac{(\alpha+\beta)}{2} \qquad (26)$$

where* $k=1/\lambda$, $\theta=(2\theta)/2$, $\alpha=\omega$, $\beta=2\theta-\omega$ or $\alpha+\beta=2\theta$. If $\omega\neq\theta$, that is an asymmetrical setting, $\alpha\neq\beta$, and the angle $\omega$ does not need to rotate $\theta$ to satisfy the Bragg condition as shown in Fig. 3(c); if $\omega=\theta$, a symmetrical setting, $\alpha=\beta=\theta$, and $\omega$ needs to rotate $\theta$ to satisfy the Bragg condition.

If the reverse rotations of the sample for $\omega$, $\chi$ and $\phi$ are performed, the components of the scattering vector $\boldsymbol{q}$ in the original coordinate system (i.e., before it was rotated into the Bragg condition), can be traced back as

$$\begin{pmatrix} q_x \\ q_y \\ q_z \end{pmatrix} = q \begin{pmatrix} \cos\varphi\cos\chi\cos(\theta-\omega) + \sin\varphi\sin(\theta-\omega) \\ -\sin\varphi\cos\chi\cos(\theta-\omega) + \cos\varphi\sin(\theta-\omega) \\ -\sin\chi\cos(\theta-\omega) \end{pmatrix}. \qquad (27)$$

Similar results can be found in pp. 154 (Aslanov *et al.*, 1998) and pp. 284 (Bennett, 2010). If it is an RSV in orthorhombic system, we have

$$\begin{pmatrix} H \\ K \\ L \end{pmatrix} = \begin{pmatrix} q_x/a^* \\ q_y/b^* \\ q_z/c^* \end{pmatrix}, \qquad (28)$$

---

* Note that these angles of $\alpha$ and $\beta$ are unrelated to those defined in the rotation matrix of Eqs. (4) or those in the conventional symbols for crystal lattice parameters.

where $a^*$, $b^*$ & $c^*$ are the lattice parameters in reciprocal space. For other crystal system lower than orthorhombic system, **B** matrix should be used to convert it from Eq. (27) to Eq. (28) (Aslanov *et al.*, 1998) (Bennett, 2010) (Busing & Levy, 1967). Set $k=1$ and $\omega = \theta$ in Eq. (27), it is just the coordinates obtained for the case of symmetrical setting as commonly used in four-circle diffractometers.

**4.2. X-ray beam condition and resolution in RSV measurement**

A typical experimental condition is as listed in Table 1 for the Diffraction Station (Beamline BL14B1) at Shanghai Synchrotron Radiation Facility (SSRF). A substrate of LaSrAlO$_4$ (LSAO) was tested for this purpose. Δα, Δχ were given from the full width at half maximum (FWHM) of rocking scans measured respectively for vertical and horizontal divergence of incident beam. Δβ was given from a $2\theta$ scan. They are combined widths, as a result of convolution of instrumental widths with the crystal diffraction widths of LSAO. As the intrinsic diffraction width of LSAO is much smaller (about arc seconds), these combined widths can be used to represent the divergences of the incident X-ray beam.

The worst errors of measuring such peak positions for general RSVs, denoted as δα, δχ and δβ respectively, are estimated as $|δα| \leq |±Δα/2|$, $|δχ| \leq |±Δχ/2|$ and $|δβ| \leq |±Δβ/2|$ respectively, which serve as a kind of accuracy of the peak positions. Furthermore, we adopt the half of δα, δχ and δβ respectively as an estimated standard deviation (ESD), to describe the margin errors of the measurement, which are much bigger than the angular precisions of the circles in the diffractometer.

Differentiating Eq. (27), the deviation of the scattering vector or RSV can be derived. For RSV (200), symmetrical setting, χ=0 ° and ϕ=0 °, $\omega=\theta$, it is worked out as

$$\delta \boldsymbol{q} = \begin{pmatrix} \delta q_x \\ \delta q_y \\ \delta q_z \end{pmatrix} = \delta q \cdot \begin{pmatrix} 1 \\ 0 \\ 0 \end{pmatrix} + q \cdot \begin{pmatrix} 0 \\ -\delta \phi \\ -\delta \chi \end{pmatrix} = \delta q \cdot \begin{pmatrix} 1 \\ 0 \\ 0 \end{pmatrix} + q \cdot \begin{pmatrix} 0 \\ -\delta \alpha \\ -\delta \chi \end{pmatrix}. \tag{29}$$

Apparently direct rocking angles α (or ϕ now) and χ measures the divergences Δα and Δχ respectively. δϕ=δα in this case.

For RSV (*013*), asymmetrical setting, χ=-90 ° and keep ϕ un-rotated at ϕ=0 °, $\chi_0$=90 °-($\theta$-$\omega$), it is a kind of ϕ-fixed case (mode 3 in the SPEC). The deviation is worked out as

$$\delta \boldsymbol{q} = \begin{pmatrix} \delta q_x \\ \delta q_y \\ \delta q_z \end{pmatrix} = \delta q \cdot \begin{pmatrix} 0 \\ \cos \chi_0 \\ \sin \chi_0 \end{pmatrix} + q \cdot \begin{pmatrix} \delta \phi \cdot \cos \chi_0 + \delta \chi \cdot \sin \chi_0 \\ \delta(\theta - \omega) \cdot \sin \chi_0 \\ -\delta(\theta - \omega) \cdot \cos \chi_0 \end{pmatrix}. \tag{30}$$

The deviations in *H, K & L* are expressed as $\begin{pmatrix} \delta H \\ \delta K \\ \delta L \end{pmatrix} = \begin{pmatrix} q\delta_x/a^* \\ \delta q_y/b^* \\ q\delta_z/c^* \end{pmatrix}.$ (31)

In above Eqs., $q$ is as shown in Eq. (26), $\delta q=2k\cdot cos\theta \cdot (\delta\theta)$, $\delta\theta=(\delta\alpha+ \delta\beta)/2$, $\delta\omega=\delta\alpha$ and $\delta\phi=\delta\chi$ in the asymmetrical case. For STO ($a$=3.9053 Å), the worst errors for measuring (*013*) in mode 3 as described above can be calculated to be less than 0.001 averagely for its *H, K & L* using Eq. (31). We choose its half-value, i.e., 0.0005 to calculate ESDs for the lattice parameters. As to STO {*200*} type, it is even less owing to the simpler and symmetrical diffraction condition. In a real measurement, it is all below 0.0005 in *H, K & L* from our experiences in the work. So this estimation should be reasonable and reliable.

In the above example of cubic ITO film, the lattice parameter can be determined in error of 0.0006 Å which is estimated from the projection error Eq. (20) of raw cell to the Niggli-reduced cell. The lattice parameters are even better determined than other films from its accurate measurement, owing to its sharp and strong film diffraction peaks.

## 5. Structural study of ferro-electric films

In this section, two examples on the determination of crystal lattice types and the lattice parameters of ferro-electric films are shown. One is single crystalline and the other is twinning.

### 5.1. Crystal system and lattice parameter determination for a PbZr$_{0.52}$Ti$_{0.48}$O$_3$ (PZT 52/48) film

The RSVs of a PZT 52/48 film on STO substrate were measured at SSLS and at SSRF, both showing very close measured RSVs as below. SRO layer as a bottom-electrode was growing between the film and substrate. As such PZT film is near the MPB composition, its structure is puzzling between monoclinic and tetragonal symmetry.

The RSMs in Fig. 4 show that a single centred spot is formed respectively for all RSVs, indicating no twin existing in film and substrate RSV spots. For STO substrate (*002*), ($\bar{1}03$) and (*013*), they were measured in 3D reciprocal space at SSRF as

  *(0.0004  -0.0011  2.0001)*
  *(-1.0005  0.0011  3.0001)*
  *(0.0034  0.9986  3.0004).*

From above set of RSVs, raw unit cell of the substrate can be derived with very good precision that $a$=3.9005 Å, $b$=3.9043 Å, $c$=3.9048 Å & $\alpha$=89.984 °, $\beta$=90.015 °, $\gamma$=90.007 °.

For PZT film, corresponding RSVs were measured at SSRF as
  *(0.0095  -0.0011  1.9010)*
  *(-0.9422  0.0001  2.8544)*
  *(0.0154  0.9578  2.8504).*

A raw cell was calculated for PZT film that $a$=4.0873 Å, $b$=4.0710 Å, $c$=4.1085 Å & $\alpha$=89.917 °, $\beta$=90.097 °, $\gamma$=89.963 °, with the same correction made as the substrate lattice parameters to the reference, i.e., cubic $a$=3.9053 Å. A tetragonal cell was finally deduced that $a$=4.079±0.008 (Å) and $c$=4.109±0.002 (Å), where the error in $a$ is calculated using the difference from the projected cell and that in $c$ is estimated from the accuracy in measuring RSVs as discussed in above sections.

To confirm such tetragonal system and lattice parameters, symmetry test has been conducted. The diffraction data and derived structure factors are as listed in Table 2. $2\theta$ has very close values for the four diffraction vectors of {*103*} family. Although the intensity correction made for the sample (irregular film shape) was not perfect, it can be seen that the deduced structure factors are close for the family. So the crystal lattice of PZT film has tetragonal symmetry with the parameters as shown above.

### 5.2. Crystal system and lattice parameter determination for a BFO film

As shown in above PZT film, RSV method can be used for a single crystalline epitaxial film to derive its crystal system and lattice parameters. For twinning film, it is also possible to obtain the crystal structure parameters if the twinning variants can be sorted out from each other to form one consistent set of RSVs. Fig. 5 shows the RSMs of a BFO film on LaSrAlO$_4$ (LSAO) substrate, grown by PLD (Pulsed Laser Deposition) (Chen, Luo, *et al.*, 2011, Chen, Luo, *et al.*, 2010, Chen, Prosandeev, *et al.*, 2011). There are four phases coexisting in the film, as shown in Fig. 5(*a*), i.e., bulk rhombohedral-like phase (marked as R-like), tetragonal-like monoclinic M$_C$ phase (as T-like, M$_C$), tilted rhombohedral-like phase (as Tri-1, 1 and 2) and tilted tetragonal-like phases (as Tri-2, I and II). All phases are also indicated in AFM topograph of Fig. 5(*d*).

The RSVs for the substrate (denoted as LSAO in the mappings) were measured to be

*(-0.0003  0.00034  1.9999)*

*(-1.0001  0.00041  3.0000)*

*(-0.00027  0.99983  2.9999).*

From above set of RSVs, raw unit cell parameters of the substrate can be obtained with very good precision that $a$=3.7577 Å, $b$=3.7590 Å, $c$=12.637 Å & $\alpha$=90.00 °, $\beta$=89.99 °, $\gamma$=90.02 °. Such lattice parameters should be corrected to its standard tetragonal ones that $a$=3.7564 Å & $c$=12.636 Å. To work out the lattice parameters for Tri-1 phase in BFO film, with selecting only one set of the twinning variants out of the eight variants, that are marked as "1" in Fig. 5(*a*), "1" in Fig. 5(*b*) and "1a" in Fig. 5(*c*), corresponding RSVs were measured to be

*(-0.0881  -0.0084  2.0170)*

*(-1.0872  -0.0061  2.9680)*

*(-0.1371  0.9710  3.0463).*

Careful identification was given to confirm and measure the above RSVs, as the spots of this set of tilted rhombohedral-like phase are not located in any coordinate plane of the reciprocal space (in the substrate coordinate system). It can only be seen a trace-like spot in Fig. 5(*a*) and 5(*b*), marked as "1" or "2". So do the spots of Tri-2 phase, marked as "I" or "II".

A raw cell for Tri-1 phase was calculated that *a*=3.9265 Å, *b*=3.8163 Å, *c*=4.1717 Å & $\alpha$=90.817 °, $\beta$=90.297 °, $\gamma$=89.359 °, with the same correction made as the substrate lattice parameters set to its standard tetragonal one. A triclinic cell was finally obtained that *a*=3.927(3) Å, *b*=3.816(3) Å, *c*=4.172(1) Å & $\alpha$=89.18(4) °, $\beta$=89.70(4) °, $\gamma$=89.36(6) °, where the errors in the parentheses were estimated from the deviations in measuring RSVs as discussed in above sections.

Tri-2 phase can be calculated to be triclinic, R-like phase to be monoclinic $M_A$ and T-like phase to be monoclinic $M_C$.

A similar example for a BFO film on LAO substrate has been investigated (Chen, Prosandeev, *et al.*, 2011), from which triclinic cells for the Tri-1 & Tri-2 phases, monoclinic $M_A$ cell for the R-like phase and monoclinic $M_C$ cell for the T-like phase have been concluded.

## 6. Discussion and summary

There is often coexistence of multiple Martersites-like twin variants in metallic alloys, intermetallics or oxide films. The distortion of cell might be very large and there are possibly overlapping peaks. In order to get one set of correct RSVs, one should carefully recognize and separate the peaks, with high resolution diffractometry set-up around intense synchrotron sources. The example in Fig. 5 shows that one set of RSVs for every morphotropic phase in BFO film has been separated and the lattice parameters can finally be determined. With separated peak intensities, one could obtain the fractional ratio of twin variants, e.g., for the abundance of *a*-domains and *c*-domains in PZT films (Lee & Baik, 1999, Nagarajan *et al.*, 1999). Otherwise, a kind of averaged structure is obtained if the peaks cannot be resolved. For such crystal structure, it could still be solved and refined using some of suitable twin laws as in normal crystal structure analysis procedure (Sheldrick, 2008), from which the variant ratio could als be worked out.

In summary, we have developed the procedure, based on RSVs, to determine Bravais lattice type and the lattice parameters for an epitaxial film. Three independent (non-coplanar) reciprocal vectors (*00L*), (*H0L*) and (*0KL*), are firstly obtained and corrected using the substrate as a reference. Three shortest vectors are then deduced to form the Niggli-reduced cell. Bravais lattice type is finally determined and its lattice parameters are accordingly calculated. Such procedure could be performed by converting or selecting the corresponding vectors in real space as well. An error in 0.001 Å or better could be reached. Some structures of multi-ferroic films have been successfully determined using the RSV method (Chukka *et al.*, 2011, Chen, Ren, *et al.*, 2011, Chen, Luo, *et al.*, 2011, Chen, Luo, *et al.*, 2010, Chen, Prosandeev, *et al.*, 2011, Chen, You, *et al.*, 2010, Chen *et al.*, 2012, Kumar *et al.*, 2013, Liu *et al.*, 2012, Liu *et al.*, 2011, Liu *et al.*, 2010, Liu *et al.*, 2013, Saito *et al.*, 2006).

The RSV method has following advantages:

1) It is a concise and direct method calculating the lattice parameters from the three reciprocal space vectors or a 6D vector in $G^6$-space, without any prior knowledge or assumption of its structure. However, in other methods, more diffraction data and iteration are needed for indexing and least-square refinement as described in Fig. 1. Hence, experimental duration for RSV method is shortened to save beam time;

2) It is an accurate method using the substrate as a reference, indpendent of X-ray wavelength and counted in instrument mis-alignment. In other methods, the wavelength should necessarily be calibrated if synchrotron X-rays are used. Although the *d*-values can be more accurately obtained using the Bragg equation with more accurate Bragg angles $\theta_B$, for example, in Bond method, there is still a difficulty in determination of the lattice type: guess and check method has to be used. Morever, such methods could cause inaccuracy in determination of the lattice parameters, especially in monoclinic angles when they are close to 90 °, particularly when the film diffraction intensity is lower, as discussed in the introduction of this article. For more discussion refer to (De Caro & Tapfer, 1998, Shilo *et al.*, 2001 , KARMAZIN & JAMES, 1972, Fatemi, 2005);

3) It is to offer not only lattice parameters but also basis vectors *a*, *b* and *c* in real space in the framework of the substrate coordiate system, i.e., the orientations of crystal bases for both the film and substate. Analyses on the length and orientation for the bases of film relative to that for substrate, will yield the information of lattice-mis-match (strain status), crystallographic tilt, step-bunching in the surface terrace (Kim *et al.*, 2011). It is worthy to mention that such bases orientation analyses may lead to an understanding for interface structure formation between film and substrate. For example, a rotation of the lattice network about the normal of the surface will result in a twist boundary between film and substrate; rotation about axis in plane in a small-angle boundary in the interface. Further exploration of the RSV method to interface study is under proposing.

**Acknowledgements**

The authors are grateful for technical support received from beamline BL14B1 of SSRF for the data collection under project No. 11sr0395 and j10sr0092. PY is supported from SSLS via NUS Core Support C-380-003-003-001. PY acknowledges the support from the Alexander von Humboldt Foundation under the ID 1031847. JW acknowledges the grant support of MOE, Singapore Ministry of Education Academic Research Fund Tier 1 (Grant Number: T11-0702-P06, R-284-000-054-112). We would thank Prof. Tom Wu and Dr. Rami Chukka for the samples.

**References**

Andrews, L. C. & Bernstein, H. J. (1988). *Acta Cryst.* **A44**, 1009-1018.

APEX2 (2009). *(version 2009.11-0), Program for Bruker CCD X-ray Diffractometer Control, Bruker AXS Inc., Madison, WI,* .

Aslanov, L. A., Fetisov, G. V. & Howard, J. A. K. (1998). *Crystallographic Instrumentation*. International Union of Crystallography.

Bennett, D. W. (2010). *Understanding Single-Crystal X-Ray Crystallography*. John Wiley & Sons.

Bond, W. L. (1960). *Acta Cryst.* **13**, 814-818.

Busing, W. R. & Levy, H. A. (1967). *Acta Crystallographica* **22**, 457-464.

Cao, W. & Cross, L. E. (1993). *Physical Review B* **47**, 4825.

Certified Scientific Software (2008). *SPEC: X-Ray Diffraction Software.*

Chen, W., Ren, W., You, L., Yang, Y., Chen, Z., Qi, Y., Zou, X., Wang, J., Sritharan, T., Yang, P., Bellaiche, L. & Chen, L. (2011). *Applied Physics Letters* **99**, 222904.

Chen, Z., Luo, Z., Huang, C., Qi, Y., Yang, P., You, L., Hu, C., Wu, T., Wang, J., Gao, C., Sritharan, T. & Chen, L. (2011). *Advanced Functional Materials* **21**, 133-138.

Chen, Z., Luo, Z., Qi, Y., Yang, P., Wu, S., Huang, C., Wu, T., Wang, J., Gao, C., Sritharan, T. & Chen, L. (2010). *Applied Physics Letters* **97**, 242903.

Chen, Z., Prosandeev, S., Luo, Z. L., Ren, W., Qi, Y., Huang, C. W., You, L., Gao, C., Kornev, I. A., Wu, T., Wang, J., Yang, P., Sritharan, T., Bellaiche, L. & Chen, L. (2011). *Physical Review B* **84**, 094116.

Chen, Z., You, L., Huang, C., Qi, Y., Wang, J., Sritharan, T. & Chen, L. (2010). *Applied Physics Letters* **96**, 252903.

Chen, Z., Zou, X., Ren, W., You, L., Huang, C., Yang, Y., Yang, P., Wang, J., Sritharan, T., Bellaiche, L. & Chen, L. (2012). *Physical Review B* **86**, 235125.

Chukka, R., Cheah, J. W., Chen, Z., Yang, P., Shannigrahi, S., Wang, J. & Chen, L. (2011). *Appl. Phys. Lett.* **98**, 242902.

Dagotto, E. (2005). *Science* **309**, 257-262.

De Caro, L. & Tapfer, L. (1998). *J. Appl. Cryst. ,* **31**, 831-834.

Fatemi, M. (2005). *Acta Crystallographica Section A* **61**, 301-313.

Fuchs, D., Arac, E., Pinta, C., Schuppler, S., Schneider, R. & Lohneysen, H. v. (2008). *Physical Review B* **77**, 014434.

Giacovazzo, C., Monaco, H. L., Artioli, G., Viterbo, D., Ferraris, G., Gilli, G., Zanotti, G. & Catti, M. (2002). *Fundamentals of Crystallography*, 2nd ed. Oxford University Press.

Haeni, J. H., Irvin, P., Chang, W., Uecker, R., Reiche, P., Li, Y. L., Choudhury, S., Tian, W., Hawley, M. E., Craigo, B., Tagantsev, A. K., Pan, X. Q., Streiffer, S. K., Chen, L. Q., Kirchoefer, S. W., Levy, J. & Schlom, D. G. (2004). *Nature* **430**, 758-761.

Hahn, T. (2006). *International Tables for Crystallography: Vol. A*. Springer.

Hwang, H. Y., Iwasa, Y., Kawasaki, M., Keimer, B., Nagaosa, N. & Tokura, Y. (2012). *Nat Mater* **11**, 103-113.

Jaffe, B., Cook, W. R. & Jaffe, H. (1971). *Piezoeletric ceramics*. Academic Press.

KARMAZIN, L. & JAMES, W. J. (1972). *Acta Cryst.* **A28**, 183-187.

Kay, H. F. & Vousden, P. (1949). *Philosophical Magazine Series 7* **40**, 1019-1040.

Kim, T. H., Baek, S. H., Jang, S. Y., Yang, S. M., Chang, S. H., Song, T. K., Yoon, J.-G., Eom, C. B., Chung, J.-S. & Noh, T. W. (2011). *Appl. Phys. Letts.* **98**, 022904-022903.

Kumar, V. S., Chukka, R., Chen, Z., Yang, P. & Chen, L. (2013). *AIP Advances* **3**, 052127.

Lee, K. S. & Baik, S. (1999). *Journal of Applied Physics* **85**, 1995-1997.

Liu, H., Yang, P., Fan, Z., Kumar, A., Yao, K., Ong, K. P., Zeng, K. & Wang, J. (2013). *Physical Review B* **87**, 220101.

Liu, H., Yang, P., Yao, K., Ong, K. P., Wu, P. & Wang, J. (2012). *Adv. Funct. Mater.* **22**, 937-942.

Liu, H., Yang, P., Yao, K. & Wang, J. (2011). *Applied Physics Letters* **98**, 102902.

Liu, H. J., Yao, K., Yang, P., Du, Y. H., He, Q., Gu, Y. L., Li, X. L., Wang, S. S., Zhou, X. T. & Wang, J. (2010). *Physical Review B* **82**, 064108.

Locquet, J. P., Perret, J., Fompeyrine, J., Machler, E., Seo, J. W. & Van Tendeloo, G. (1998). *Nature* **394**, 453-456.

Ma Zhesheng & Shi Nicheng (1995). *X-ray Crystallography: basic theory and experimental technique for crystal structure determination*. Wu han Zhong guo di zhi ta xue chu ban she (in Chinese).

Nagarajan, V., Jenkins, I. G., Alpay, S. P., Li, H., Aggarwal, S., Salamanca-Riba, L., Roytburd, A. L. & Ramesh, R. (1999). *Journal of Applied Physics* **86**, 595-602.

Niggli, P. (1928). *Krystallographische und strukturtheoretische Grundbegriffe.* Akad. Verl.-Ges.

Paciorek, W. A. & Bonin, M. (1992). *Journal of Applied Crystallography* **25**, 632-637.

Pecharsky, V. K. & Zavalij, P. Y. (2003). *Fundamentals of Powder Diffraction and Structural Characterization of Materials, 2nd Ed*. Kluwer Academic Publishers.

Saito, K., Ulyanenkov, A., Grossmann, V., Ress, H., Bruegemann, L., Ohta, H., Kurosawa, T., Ueki, S. & Funakubo, H. (2006). *Jpn. J. Appl. Phys. Part 1 - Regul. Pap. Brief Commun. Rev. Pap.* **45**, 7311-7314.

Schlom, D. G., Chen, L. Q., Eom, C. B., Rabe, K. M., Streiffer, S. K. & Triscone, J.-M. (2007). *Annual Review of Materials Research* **37**, 589-626.

Schmidbauer, M., Kwasniewski, A. & Schwarzkopf, J. (2012). *Acta Crystallographica Section B* **68**, 8-14.

Sheldrick, G. (2008). *Acta Crystallographica Section A* **64**, 112-122.

Shilo, D., Lakin, E. & Zolotoyabko, E. (2001 ). *Journal of Applied Crystallography* **34**, 715-721.

Wadley, P., Crespi, A., Gazquez, J., Roldan, M. A., Garcia, P., Novak, V., Campion, R., Jungwirth, T., Rinaldi, C., Marti, X., Holy, V., Frontera, C. & Rius, J. (2013). *Journal of Applied Crystallography* **46**.

Yang, P. (October 2012). *Lecture Notes, Autumn School on X-ray diffractometry, Shanghai Synchrotron Radiation Facility (SSRF),* 369-394.

Zeches, R. J., Rossell, M. D., Zhang, J. X., Hatt, A. J., He, Q., Yang, C. H., Kumar, A., Wang, C. H., Melville, A., Adamo, C., Sheng, G., Chu, Y. H., Ihlefeld, J. F., Erni, R., Ederer, C., Gopalan, V., Chen, L. Q., Schlom, D. G., Spaldin, N. A., Martin, L. W. & Ramesh, R. (2009). *Science* **326**, 977-980.

Zubko, P., Gariglio, S., Gabay, M., Ghosez, P. & Triscone, J.-M. (2011). *Annual Review of Condensed Matter Physics* **2**, 141-165.

**Table 1**  Experimental condition of the Diffractometry Station (Beamline BL14B1) at SSRF

| | |
|---|---|
| 6-circle Huber diffractometer (core four-circle diffractometer) | |
| X-ray wavelength λ normally used: | 1.2398 Å (10 keV photons) |
| Distance from sample to scintillator detector: | 660 mm |
| Beam size normally used: | 0.300×0.300 mm$^2$ |
| Detector slits normally used: | 0.400×0.400 mm$^2$ |
| *Vertical divergence of incident beam: | Δα=0.0050° (Δω in the scattering plane) |
| *Horizontal divergence of incident beam: | Δχ=0.060 ° (perpendicular to the scattering plane) |
| *Subtended angle of detector to sample centre: | Δβ=0.019° in 2θ scan |

* Data from SSRF measurement using LSAO substrate (002) in the framework of diffractometer coordinate system.

**Table 2**  Tetragonal symmetry test for PZT 52/48 film on SRO/STO.

Photon energy: ≈10 keV (λ=1.2381 Å), Intensity data from SSRF, 19 June 2011.

| *HKL* | *2θ$_B$* | *Integrated intensity I *, a.u.* | *Structure factor |F| *, a.u.* |
|---|---|---|---|
| ($\bar{1}$03) | 56.96 ° | 1.866(7) ×10$^6$ | 1.366 (2) |
| (013) | 56.95 ° | 1.777(7) ×10$^6$ | 1.333(3) |
| (103) | 56.97 ° | 1.681(6) ×10$^6$ | 1.296 (2) |
| (0$\bar{1}$3) | 56.97 ° | 1.805(6) ×10$^6$ | 1.343 (2) |

* Numbers in the brackets are estimated standard deviations (ESD) only from counting statistics for integrated intensity and structure factor.

**Figure 1** General procedure in crystal structure determination. Diffraction peak position data can be collected either with angles *2θ, ω, χ* and *ϕ* or angle mappings for bulk materials or single crystalline epitaxial film, or even *2θ* alone for poly-crystalline films. Then indexing procedure is starting to obtain possible lattice type and its parameters, as shown in the left-half of the figure; alternatively, diffraction data can also be collected using 2-dimension (2D) reciprocal space mapping (RSM) and 3-dimension (3D) RSV with Miller indices *H, K & L* as coordinates. At least three shortest independent vectors should be obtained and served as basis vectors, as shown in the right-half of the figure. Next step, the Niggli cell is reduced and the Bravais lattice type with lattice parameters are finally worked out. Whole set of diffraction data for its structure determination can subsequently be collected and the crystal structure is solved. For the iteration methods of TREOR, ITO, DICVOL, refer to pp. 399 (Pecharsky & Zavalij, 2003).

**Figure 2** Lattice and coordinate system *H, K & L* in a 3D reciprocal space, O as the origin. ***a\****, ***b\**** & ***c\**** (not shown) are basis vectors of the reciprocal lattice along ***H, K & L*** direction respectively. ***q*** and ***$q_0$*** are RSVs pointing to a general position and a lattice point respectively.

**Figure 3** Three orthogonal and right-handed coordinate systems for a standard four-circle diffractometer with Euler angles *2θ, ω, χ* and ϕ. The coordinate system convention is as proposed in the SPEC (Certified Scientific Software, 2008). When the coordinates are all zero they are coincident as the same for the three coordinate systems. (*a*) A laboratory coordinate system *xyz* (fixed frame in laboratory) and a diffractometer coordinate system (angles as shown). Circles 2θ, ω, and ϕ, are defined as right-handed and χ left-handed as indicated. (*b*) A sample coordinate system (fixed with sample natural axes). A scattering vector ***q***, along the normal of the Bragg planes, is oriented with angles χ and ϕ before it is rotated into a Bragg condition. The component $q_z$ is along z-direction and $q_\phi$ in the *xy*-plane. (*c*) A Bragg condition is satisfied as shown. The scattering vector ***q*** is firstly rotated into the scattering plane, i.e., *xy*-plane in the coordinate system with angle *θ-ω* against the $q_z$ axis. With a succeeding rotation ω in the scattering plane about axis *z*, the diffraction then occurs that ***q*** = ***$k_2$*** − ***$k_1$***, (|***$k_1$***|=|***$k_2$***|=1/λ). Its magnitude *q* is equal to a reciprocal space vector – RSV, i.e., 1/*d*, reciprocal of the Bragg plane spacing. Here the incident X-ray ***$k_1$*** is at angle α (≡ω) and diffracted X-ray ***$k_2$*** at angle β respectively to the component $q_\phi$. When *ω=θ*, *α=β=θ*, it is symmetrical setting and *ω* needs to rotate *θ* to satisfy the Bragg condition.

**Figure 4** RSMs of PbZr$_{0.52}$Ti$_{0.48}$O$_3$ (PZT 52/48) film on SRO/STO, with diffraction intensities in logarithm scale and grayscale bar as shown. The data were collected at SSLS, X-ray wavelength

$\lambda$=1.5405 Å. *(a)* (*002*) *HL* mapping; *(b)* (*002*) *HK* mapping, L=1.901; *(c)* ($\bar{1}$*03*) *HL* mapping and *(d)* (*013*) *KL* mapping. SRO film looks fully strained to the substrate (straight down the STO spots in the mappings, out of our interests in this article and not paid more attention here). PZT film shows a single spot in every mapping, no sign of twinning. Note that *a* and *b* were chosen differently in the experiments at SSLS and SSRF. But the results are consistent.

**Figure 5** RSMs of BFO film on LSAO, with diffraction intensities in logarithm scale and grayscale bar as shown. The data were collected at SSRF, X-ray wavelength $\lambda$=1.2398 Å. There are four phases coexisting in the film, as shown in Fig. 5(a), i.e., bulk rhombohedral-like phase (marked as R-like), tetragonal-like monoclinic Mc phase (as T-like, Mc), tilted rhombohedral-like phase (as Tri-1, 1 and 2) and tilted tetragonal-like phases (as Tri-2, I and II). *(a)* (*002*) *HL* mapping of Tri-1 phase. Spots 1 and 2 are weaker, as the mapping was penetrating between the 1a and 1b, 2a and 2b pairs of this phase. So are the spots I and II of the Tri-2 phase; *(b)* ($\bar{1}$*03*) *HL* mapping. They are even weaker than in (*a*) as they are more apart from each other for the pairs. As a result, the position of sport 2 is estimated in the figure. (*013*) RSM shows similar sports as here, the mapping not shown; *(c)* (*002*) *HK* mapping, *L*=2.017, Tri-1 phase. There are eight twinning variants with a four-fold symmetry. There is a similar pattern for the twins of Tri-2 phase (*L*=1.795, not shown); and *(d)* AFM topograph of the film with the phases indicated. The flat area corresponds to T-like, Mc phase. The stripe-like region corresponds to the tilted rhombohedral-like phase and tilted tetragonal-like phase, i.e., mixture of Tri-1 phase and Tri-2 phase, which is similar to the phases in the BFO film on LAO (Chen, Prosandeev*, et al.*, 2011). Bulk R-like phase may be hiding in some gaps between the tilted phases.

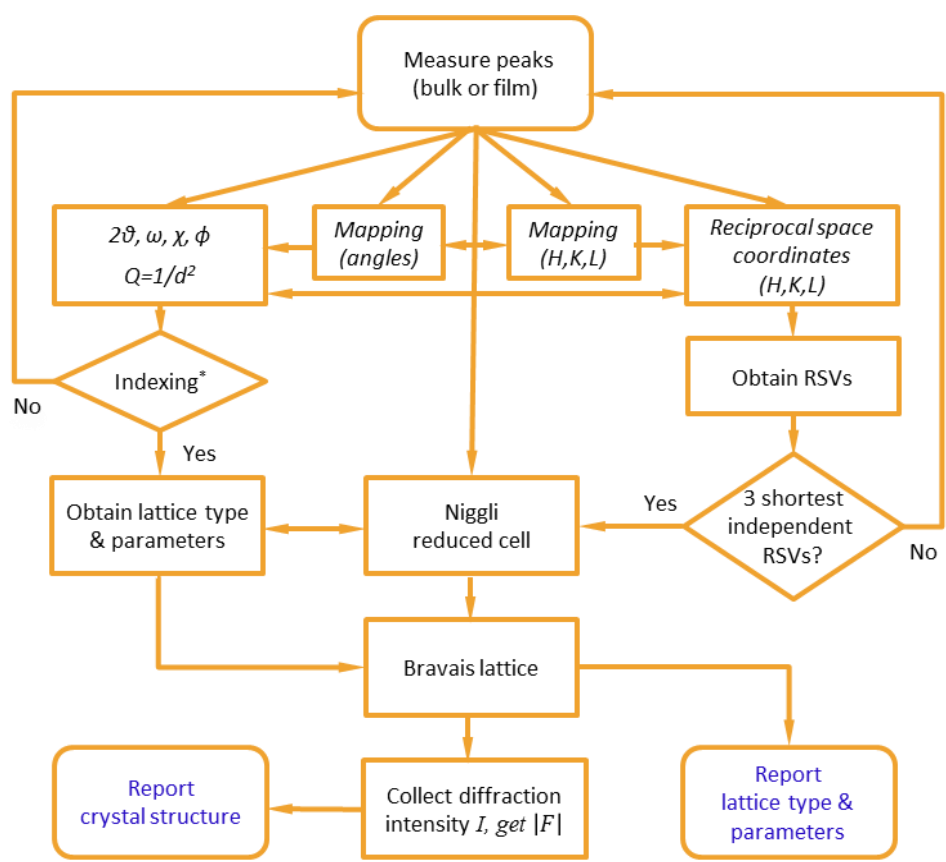

**Figure 1** General procedure in crystal structure determination. Diffraction peak position data can be collected either with angles *2θ, ω, χ* and *ϕ* or angle mappings for bulk materials or single crystalline epitaxial film, or even *2θ* alone for poly-crystalline films. Then indexing procedure is starting to obtain possible lattice type and its parameters, as shown in the left-half of the figure; alternatively, diffraction data can also be collected using 2-dimension (2D) reciprocal space mapping (RSM) and 3-dimension (3D) RSV with Miller indices *H*, *K* & *L* as coordinates. At least three shortest independent vectors should be obtained and served as basis vectors, as shown in the right-half of the figure. Next step, the Niggli cell is reduced and the Bravais lattice type with lattice parameters are finally worked out. Whole set of diffraction data for its structure determination can subsequently be collected and the crystal structure is solved. For the iteration methods of TREOR, ITO, DICVOL, refer to pp. 399 (Pecharsky & Zavalij, 2003).

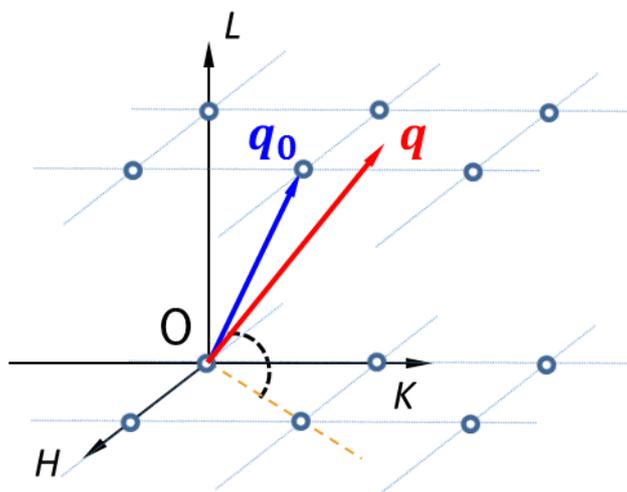

**Figure 2** Lattice and coordinate system *H, K* & *L* in a 3D reciprocal space, O as the origin. ***a\**, *b\** & *c\**** (not shown) are basis vectors of the reciprocal lattice along ***H, K & L*** direction respectively. ***q*** and ***q*$_0$** are RSVs pointing to a general position and a lattice point respectively.

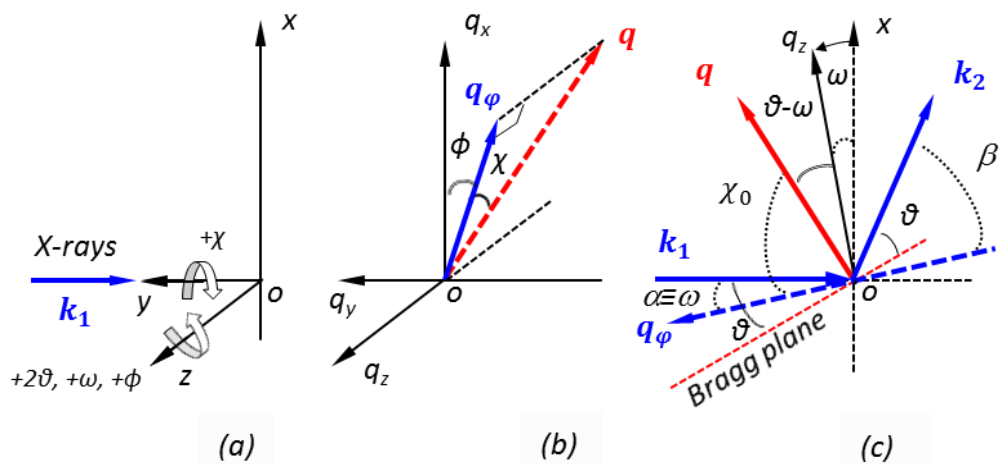

**Figure 3** Three orthogonal and right-handed coordinate systems for a standard four-circle diffractometer with Euler angles *2θ, ω, χ* and ϕ. The coordinate system convention is as proposed in the SPEC (Certified Scientific Software, 2008). When the coordinates are all zero they are coincident as the same for the three coordinate systems. (*a*) A laboratory coordinate system *xyz* (fixed frame in laboratory) and a diffractometer coordinate system (angles as shown). Circles 2θ, ω, and ϕ, are defined as right-handed and χ left-handed as indicated. (*b*) A sample coordinate system (fixed with sample natural axes). A scattering vector *q*, along the normal of the Bragg planes, is oriented with angles χ and ϕ before it is rotated into a Bragg condition. The component $q_z$ is along z-direction and $q_\phi$ in the *xy*-plane. (*c*) A Bragg condition is satisfied as shown. The scattering vector *q* is firstly rotated into the scattering plane, i.e., *xy*-plane in the coordinate system with angle *θ-ω* against the $q_z$ axis. With a succeeding rotation ω in the scattering plane about axis *z*, the diffraction then occurs that $\boldsymbol{q} = \boldsymbol{k}_2 - \boldsymbol{k}_1$, ($|\boldsymbol{k}_1|=|\boldsymbol{k}_2|=1/\lambda$). Its magnitude *q* is equal to a reciprocal space vector – RSV, i.e., 1/*d*, reciprocal of the Bragg plane spacing. Here the incident X-ray $\boldsymbol{k}_1$ is at angle α (≡ω) and diffracted X-ray $\boldsymbol{k}_2$ at angle β respectively to the component $q_\phi$. When *ω=θ*, α=β=*θ*, it is symmetrical setting and *ω* needs to rotate *θ* to satisfy the Bragg condition.

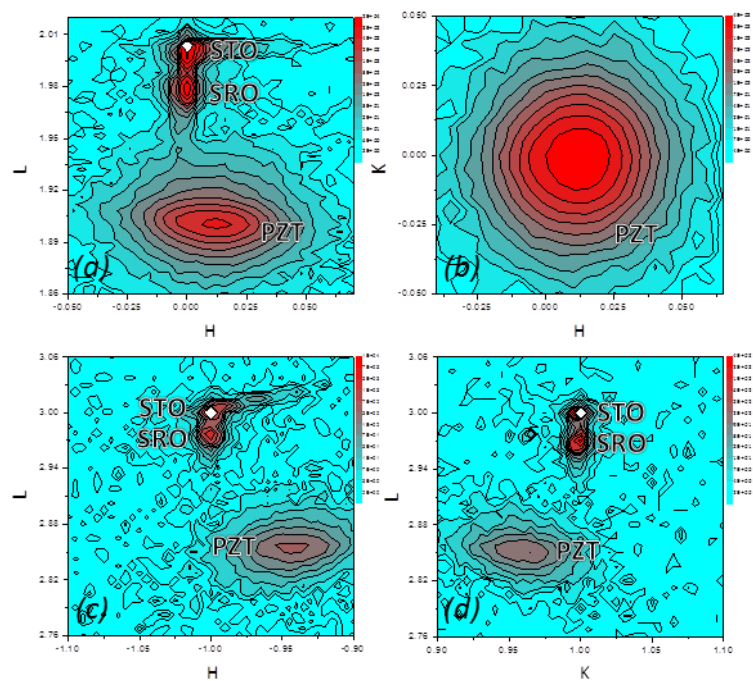

**Figure 4** RSMs of PbZr$_{0.52}$Ti$_{0.48}$O$_3$ (PZT 52/48) film on SRO/STO, with diffraction intensities in logarithm scale and grayscale bar as shown. The data were collected at SSLS, X-ray wavelength λ=1.5405 Å. *(a)* (*002*) *HL* mapping; *(b)* (*002*) *HK* mapping, L=1.901; *(c)* ($\bar{1}$*03*) *HL* mapping and *(d)* (*013*) *KL* mapping. SRO film looks fully strained to the substrate (straight down the STO spots in the mappings, out of our interests in this article and not paid more attention here). PZT film shows a single spot in every mapping, no sign of twinning. Note that *a* and *b* were chosen differently in the experiments at SSLS and SSRF. But the results are consistent.

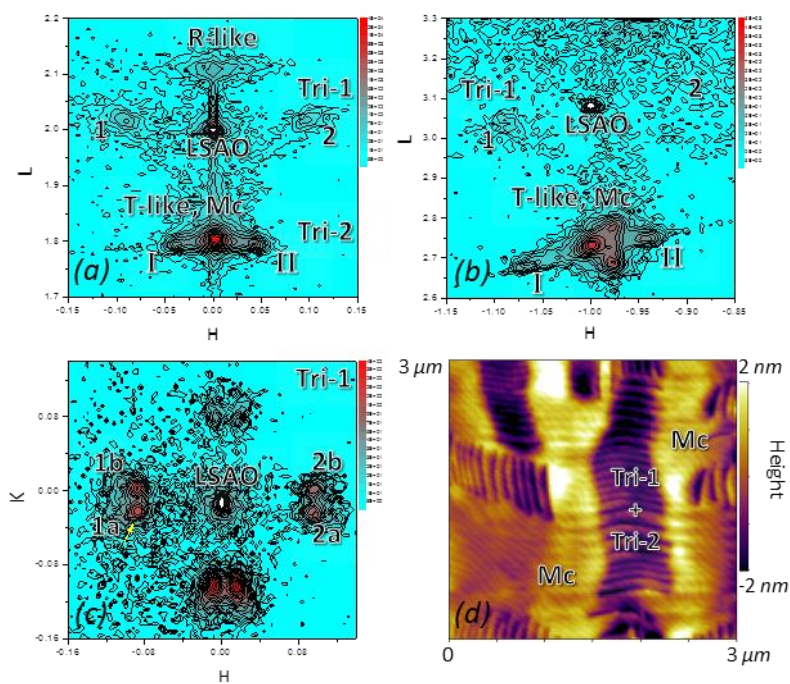

**Figure 5** RSMs of BFO film on LSAO, with diffraction intensities in logarithm scale and grayscale bar as shown. The data were collected at SSRF, X-ray wavelength λ=1.2398 Å. There are four phases coexisting in the film, as shown in Fig. 5(a), i.e., bulk rhombohedral-like phase (marked as R-like), tetragonal-like monoclinic Mc phase (as T-like, Mc), tilted rhombohedral-like phase (as Tri-1, 1 and 2) and tilted tetragonal-like phases (as Tri-2, I and II). *(a)* (*002*) *HL* mapping of Tri-1 phase. Spots 1 and 2 are weaker, as the mapping was penetrating between the 1a and 1b, 2a and 2b pairs of this phase. So are the spots I and II of the Tri-2 phase; *(b)* ($\bar{1}03$) *HL* mapping. They are even weaker than in (*a*) as they are more apart from each other for the pairs. As a result, the position of sport 2 is estimated in the figure. (*013*) RSM shows similar sports as here, the mapping not shown; *(c)* (*002*) *HK* mapping, $L$=2.017, Tri-1 phase. There are eight twinning variants with a four-fold symmetry. There is a similar pattern for the twins of Tri-2 phase ($L$=1.795, not shown); and *(d)* AFM topograph of the film with the phases indicated. The flat area corresponds to T-like, Mc phase. The stripe-like region corresponds to the tilted rhombohedral-like phase and tilted tetragonal-like phase, i.e., mixture of Tri-1 phase and Tri-2 phase, which is similar to the phases in the BFO film on LAO (Chen, Prosandeev*, et al.*, 2011). Bulk R-like phase may be hiding in some gaps between the tilted phases.